
%

\input phyzzx
\tolerance=1000
\sequentialequations
\def\rl{\rightline}

\def\t1{{\tilde 1}}

\def\PRL#1#2#3{Phys. Rev. Lett. {\bf#1} (19#2) #3}

\REF\BH{R. Barbieri and L.J. Hall, \PRL{68}{92}{752}.}
\REF\REST{The ALEPH collaboration, CERN--PPE/92--23, February 1992;
The L3 collaboration, CERN--PPE/92--58; The DELPHI collaboration, Z. Phys
C54 (1992) 55.}
\REF\OPAL{The OPAL collaboration, CERN--PPE/92--18, February 1992.}
\REF\HEBEKER{T. Hebbeker, PITHA 91/17, October 1991.}
\REF\SUM{The LEP collaborations, ALEPH, DELPHI, L3, OPAL,
CERN--PPE/91--232.}

\singlespace
\rl{SSCL--Preprint--126}
\rl{WIS--92/50/JUN--PH}
\rl{June 17, 1992; revised August 14, 1992}
\normalspace
\medskip
\titlestyle{\bf Comment On ``Grand Unification and Supersymmetric Threshold"}
\author{Alon E. Faraggi
{\footnote*{email: fhalon@weizmann.bitnet}}}
\centerline {Department of Physics, Weizmann Institute of Science}
\centerline {Rehovot 76100, Israel}
\author{Benjam\'\i n Grinstein\footnote\dagger{On leave of absence from Harvard
 University. email: grinstein@sscvx1.ssc.gov}\ and
Sydney Meshkov\footnote\ddagger{email: syd@sscvx1.ssc.gov}}
\centerline{Superconducting Super Collider Laboratory}
\centerline{2550 Beckleymeade Ave.}
\centerline{Dallas, TX 75237}
\bigskip\bigskip
\centerline{Submitted to {\it Physical Review Letters}}
\bigskip
\centerline{PACS numbers: 12.10.Dm, 11.30.Pb, 14.80.Ly}
\nopagenumbers
\pageno=0
\singlespace
\vskip 0.5cm
\endpage
\normalspace

\pagenumbers
In a recent letter, Barbieri and Hall [\BH] consider the effects of ultra
heavy particle thresholds on the unification of gauge coupling constants
in the minimal supersymmetric unified theory based on $SU(5)$. To test
for unification of the gauge couplings $\alpha_i(\mu)$, $i=1,2,3$ at some
 scale
$\mu=M_V$, one extrapolates the measured values $\alpha_i=\alpha_i(M_Z)$
to the supersymmetric threshold, $\mu=M_S$. At that point the rules for
extrapolation ---the $\beta$ functions--- change because of additional
 degrees
of freedom that become active. Normally one extrapolates further to
$\mu\gg M_S$ and tests whether there is a scale $\mu=M_V$, such that
$\alpha_1(M_V)=\alpha_2(M_V)=\alpha_3(M_V)$. One can thus determine
whether a SUSY scale $M_S$ exists such that unification occurs;
this, in turn, {\it determines} $M_S$.

Barbieri and Hall point out that this normal procedure implicitly
assumes that all of the ultra-heavy particles ---those whose masses are
of order $M_V$--- are degenerate. Relaxing this artificial assumption
gives additional modifications to the extrapolation rules. For every new
ultra heavy particle there is a corresponding modification to the
extrapolation rules. This introduces additional unknown parameters.
As a result one can achieve unification for a wide range of $M_S$.
One can no longer determine the threshold for supersymmetric particles
at all. Barbieri and Hall summarize their result in the following
equation\footnote{*}{The coefficients of the logarithms differ from those of
Barbieri and Hall. This does not affect their conclusions.}:
$$\ln{{M_S}\over{M_Z}}=I+{5\over{22-3n_h}} \ln{{M_V}\over{M_{\Sigma}}}-
{13\over{22-3n_h}}\ln{{M_V}\over{M_H}}, \eqno(1)$$
where $M_H$, $M_{\Sigma}$ and $M_V$ are the masses of the super multiplets
containing the colored triplet higgs, the adjoint higgs, and the
super heavy vector bosons, respectively,  $n_h$ is the number of higgs doublets
lighter than $M_S$, and $I$ is given in terms of
measured inputs:
$$I={{4\pi}\over{(22-3n_h)\alpha}}[3-15\sin^2\theta_W+7{\alpha\over\alpha_s}].
\eqno(2)$$
For $n_h=1$, $\alpha^{-1}=127.8\pm0.2$,
$\sin^2\theta_W=0.2334\pm0.0008$ and  $\alpha_s=0.115\pm0.007$,
they find $I=-2.1\pm2.6\pm1.0$.

The precision of the determination of $M_S$ depends crucially on how
distinct from
$M_V\sim10^{16}GeV$ the other two free parameters
$M_\Sigma$ and $M_H$ can be. Barbieri and Hall use $M_H\geq10^{10}GeV$.
With this lower bound and assuming $M_H$ is
smaller than the Planck mass, one gets
$$-9.5<-{{13}\over{19}}\ln{{M_V}\over{M_H}}<4.7 \eqno(3)$$
Together with eq.~(1) this gives bounds on $M_S$ that
are not at all restrictive.

We point out that there are some quite solid theoretical and experimental
constraints on the ultra heavy masses that were overlooked in Ref. [\BH].
Proton stability gives a rather stringent bound $M_H>x\cdot10^{16}GeV$,
where $x$ depends on the details of the model but is of order of unity. This
is readily estimated as follows. Exchange of a colored fermion in the
$H$--$\bar H$
 multiplets gives rise to B and L violating, dimension five operators
(squark--squark--quark--lepton), with coefficient
${\lambda^2_q}/{M_H}$, where
$\lambda_q$ is the light--quark Yukawa
coupling. Gluino exchange yields  $B$ and $L$ violating operators
of dimension 6 involving light quarks and leptons only, with coefficient
$(\lambda_q^2/M_H)(\alpha_3/\pi)1/M_S$.
For proton stability this must be smaller than $10^{-30}GeV^{-2}$,
giving $M_H>10^{16}GeV$.

An upper bound on $M_H$ can be obtained by requiring internal
consistency. The most general superpotential for the
$\Sigma$, $H$, ${\bar H}$ fields is
$$W=m{\bar H}H+\lambda{\bar H}\Sigma{H}+{1\over2}
{\lambda^\prime}vTr\Sigma^2+{1\over3}{\lambda^\prime}Tr \Sigma^3
\eqno(4)$$
and $\langle\Sigma\rangle=v$diag$(2,2,2,-3,-3)$, where
$v={{M_V}/{g_{_U}}}$. The Higgs doublet will be massless only if
$m$ is fine tuned. Thus $M_H$ is determined by $\lambda$ and
$\lambda^\prime$ (and $M_V$). The requirement that there be no Landau
poles in these dimensionless couplings below the scalar mass $M_H$
gives an upper bound $M_H$ of only a few times $v$ (the precise number
requires non--perturbative calculations, beyond the scope of this
commentary). Since $g_{_U}\sim0.7$, one has, conservatively,
$M_H\leq10M_V$; thus,
$\vert{{13}\over{19}}\ln{{M_V}\over{M_H}}\vert\leq1.6$. Since $M_H$ cannot
differ from $M_V$ by more than an order of
magnitude the leading log calculation of threshold effects
leading to eq.~(1) is inadequate: and a full one loop analysis -- beyond
the scope of this {\it comment} --- is needed.

While a triviality argument
again gives an upper bound on $M_\Sigma$, no useful
lower bound  can be
obtained from proton stability considerations.  Recent LEP results
suggest a larger value for $\alpha_s$ than the one quoted above.
With[\REST,\OPAL]
$\alpha_s=0.125\pm0.005$  and  $\alpha$ and
$\sin^2\!\theta_W$ as above, one has $I=-5.3\pm2.5$ or
$I<-2.8(1\sigma)$ (here and below we specialize to the $n_h=2$ case).
Neglecting heavy thresholds, this gives $M_S\leq1.1M_Z$,
a somewhat uncomfortably low value. The bound is
weakened if $M_\Sigma$ is substantially smaller than $M_V$. However,
as $M_\Sigma/M_V$ decreases, so does $M_Z/M_V$, so that eventually $M_V$
reaches the Planck scale:
$$
\ln{M_Z\over M_V} = \tilde I +{2\over3}\ln {M_Z\over M_S}
\qquad\hbox{and}\qquad
\ln{M_Z\over M_\Sigma} = \hat I -{38\over15} \ln{M_Z\over M_S},\eqno(5a,b)$$
where
$$
\tilde I = {\pi\over6\alpha}[ 3 -18\sin^2\theta_W+ 10{\alpha\over\alpha_s}]
\qquad\hbox{and}\qquad
\hat I = {\pi\over30\alpha}[ -57 +270\sin^2\theta_W -118{\alpha\over\alpha_s}]
\eqno(6a,b)
$$
For $M_S=M_Z$ ($M_S=10M_Z$) and $\alpha$ and $\sin^2\theta_W$ as above, this
occurs for $\alpha_s=0.128$ ($\alpha_s=0.123$ )
and $M_\Sigma=8.1\times10^8$GeV ($M_\Sigma=1.2\times10^8$GeV ), where
$M_{Planck}=1.2\times10^{19}$.

Larger values of $\alpha_s$ can be
accommodated by breaking  the degeneracy of the
$\Sigma$-multiplet.
Under $SU(2)\times SU(3)$ this contains a singlet, a $(3,1)$ of
mass $M_3$ and a $(1,8)$ of mass $M_8$. For $M_3<M_8$ one has
$$
\ln{M_Z\over M_V} = \tilde I +{2\over3}\ln {M_Z\over M_S} -
{25\over12}\ln{M_3\over M_8}
\qquad\hbox{and}\qquad
\ln{M_Z\over M_3} = \hat I -{38\over15} \ln{M_Z\over M_S}+{{59}\over{12}}
\ln{{M_3}\over{M_8}},\eqno(7a,b)
$$
so that unification is attained for $M_V<M_{\rm Planck}$, with $M_S/M_Z$
undetermined. In minimal SUSY $SU(5)$-GUT, $M_3=M_8$ ({\it cf.} eq.~(4)).
However, $M_3\ne M_8$ may be induced by higher dimension operators.
Since $M_\Sigma/M_V\ll M_V/M_{\rm Planck}$, the condition $M_3\ll M_8$
may be attained easily.

Two different methods are used at LEP to extract
$\alpha_s$ [\HEBEKER,\SUM]. The ``event topology" method has smaller
error bars, which are mainly theoretical. The second method involves
measurement of the ratio ${{\Gamma_{had}}/{\Gamma_{lep}}}$ at the
$Z$ peak and yields
$\alpha_s=0.148\pm0.021$ [\OPAL].
If the estimate of theoretical uncertainties in the first method proves
to be too optimistic, and if $\alpha_s$ turns out to be as large as
 suggested by
the second method,
the Minimal Supersymmetric GUT Model based on $SU(5)$ may remain
compatible with experiment, but the original compelling simplicity will be
lost.

{\it Acknowledgments.}
We thank Steve Kelley,
Bryan W. Lynn, Patricia McBride and Makoto Takashima for
very useful discussions. AF would like to thank the Feinberg school
for support and the SSC laboratory for its hospitality.
BG would like to thank the Alfred P. Sloan
Foundation for partial support. This work  was supported in part
by the Department of Energy under contract DE--AC35--89ER40486.


\refout

\vfill
\eject

\end
\bye